\documentclass[twocolumn]{IEEEtran} 
\usepackage{amsmath}
\usepackage{amsfonts}   
\usepackage{amssymb}
\usepackage[final]{graphicx}

\def\BibTeX{{\rm B\kern-.05em{\sc i\kern-.025em b}\kern-.08em
    T\kern-.1667em\lower.7ex\hbox{E}\kern-.125emX}}

\newcommand\beq{\begin{equation}}
\newcommand\eeq{\end{equation}}

\usepackage[final]{graphicx}

\begin{document}

\title{Paired Cut-Wire Arrays for Enhanced Transmission of Transverse-Electric Fields through Sub-Wavelength Slits in a Thin Metallic Screen}

\author{
Ilaria Gallina, Giuseppe Castaldi, Vincenzo Galdi, {\em Senior Member, IEEE}, Emiliano Di Gennaro,\protect\\
and Antonello Andreone, {\em Member, IEEE}
\thanks{
I Gallina, G. Castaldi, and V. Galdi are with the Waves Group, Department of Engineering, University of Sannio, I-82100 Benevento, Italy (e-mail:
ilaria.gallina@unisannio.it, castaldi@unisannio.it, vgaldi@unisannio.it).\protect\\
E. Di Gennaro and A. Andreone are with the CNR-SPIN and Department of Physics, University of Naples ``Federico II,'' I-80125 Naples, Italy (e-mail:
emiliano@na.infn.it, andreone@unina.it).}
}
\markboth{GALLINA et al.: Paired Cut-Wire Arrays for Enhanced Transmission of TE Fields through Sub-Wavelength Slits}{} 
\maketitle

\date{\today}

\begin{abstract}
It has recently been shown that the transmission of electromagnetic fields through sub-wavelength slits (parallel to the electric field direction) in a thin metallic screen can be greatly enhanced by covering one side of the screen with a metallic cut-wire array laid on a dielectric layer. In this Letter, we show that a richer phenomenology (which involves both electric- and magnetic-type resonances) can be attained by pairing a second cut-wire array at the other side of the screen. Via a full-wave comprehensive parametric study, we illustrate the underlying mechanisms and explore the additional degrees of freedom endowed, as well as their possible implications in the engineering of enhanced transmission phenomena.
\end{abstract}

\maketitle 


\section{Introduction}
\PARstart{T}{riggered by} the seminal work by Ebbesen {\em et al.} \cite{Ebbesen1}, the study of extraordinary transmission phenomena through sub-wavelength apertures has emerged during the last decade as one of the most fascinating research areas in electromagnetics and optics, with a wealth of subtle physical aspects and potentially disruptive application perspectives to data storage, imaging, sensing, and lithography  technologies. The reader is referred to \cite{Weiner} and \cite{Vidal} (and references therein) for recent tutorials and reviews of the well-established physical aspects and the as yet unsettled issues. 

For the slit geometries of direct interest for the present investigation, most of the available results \cite{Weiner}, \cite{Vidal} pertain to the transverse-magnetic (TM) polarization (i.e., magnetic field directed along the slits), for which the enhanced transmission is mediated by a {\em propagating} waveguide mode (with no cut-off frequency) supported by the slits. Comparatively much less effort (see, e.g., \cite{Moreno1}--\cite{Nikitin}) has been devoted to the transverse-electric (TE) case (i.e., electric field directed along the slits), for which the slit waveguide mode does instead exhibit a nonzero cut-off frequency, below which only {\em evanescent} coupling is possible. In particular, substantial (nearly 800-fold) transmission enhancements of TE fields were observed in \cite{Jin} by placing a metallic cut-wire array (with the wires centered on the slits and parallel to them) on a thin dielectric substrate at the side of the screen that is directly illuminated. Such phenomenon was shown to be attributable to the excitation of an electric (dipole-like) resonance in the cut wires, whose strong field localization near the input aperture of the slit allows effective coupling of the illuminating plane-wave with the evanescent spectrum, and thus its ``squeezing'' through the slits \cite{Jin}. 
Building up on these results, in this Letter, we show that richer effects (and moderately higher transmission enhancements) can be attained by pairing a second cut-wire array at the other side of the screen. 

Cut-wire pairs have recently elicited attention as building blocks for (magnetic or negative-index) metamaterials (see, e.g., \cite{Linden}--\cite{Donzelli}).
For the scenario of interest, we show that such structures can induce TE enhanced transmission phenomena via a richer phenomenology, whose underlying mechanisms can be tuned, by acting on the structure parameters (see, e.g., \cite{Lee1}--\cite{Burokur}), thereby providing new degrees of freedom for engineering the enhanced transmission phenomena.

%
\begin{figure}
\begin{center}
\includegraphics[width=8.8cm]{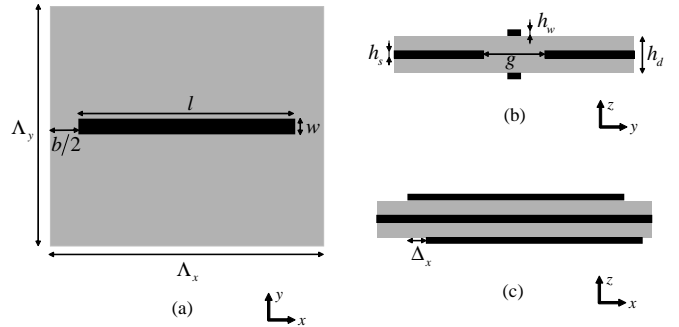}
\end{center}
\caption{Top (a) and side (b,c) views of the unit cell (details in the text), with black and gray regions representing metallic and dielectric materials, respectively. Thicknesses are exaggerated for visualization purposes.}
\label{Figure1}
\end{figure}

\section{Problem geometry}
The geometry of interest, obtained via two-dimensional (2-D) replication of the unit cell (of periods $\Lambda_x$ and $\Lambda_y$) illustrated in Fig. \ref{Figure1}, is constituted by a metallic screen with infinitely long (in the $x$ direction) slits of subwavelength width $g$ [Fig. \ref{Figure1}(b)] separated by a distance $\Lambda_y$ [Fig. \ref{Figure1}(a)]. Generalizing the scenario in \cite{Jin}, in our configuration, the screen (of thickness $h_s$) is embedded in a dielectric slab of relative permittivity $\varepsilon_r$ and thickness $h_d$ [Fig. \ref{Figure1}(b)], with a 1-D array of metallic wires (of length $l$, width $w$, and thickness $h_w$) laid at each side of the slab along the $x$ direction. The arrays are located at the center of each slit, with period $\Lambda_x$ (i.e., longitudinal gap $b=\Lambda_x-l$), and a possible longitudinal displacement $\Delta_x$ [Fig. \ref{Figure1}(c)]. 
We study the time-harmonic response, under TE-polarized ($x-$directed electric field) plane-wave excitation, via the finite-integration commercial software package CST Microwave Studio \cite{CST}. In our simulations below, we assume an aluminum (electrical conductivity $\sigma=3.72\times10^7$ S/m) screen, copper ($\sigma=5.8\times10^7$ S/m) wires, and a lossless dielectric, unless otherwise stated. The 3-D unit cell in Fig. \ref{Figure1} is terminated via periodic boundary conditions in the $x-y$ plane and with open-boundary (matched) conditions along $z$ (with a 10 mm air layer at each side of the slab), and is discretized using an adaptive hexahedral mesh (resulting, within the parametric range of interest, in about $6\times 10^4$ cells).

%
\begin{figure}
\begin{center}
\includegraphics[width=8.8cm]{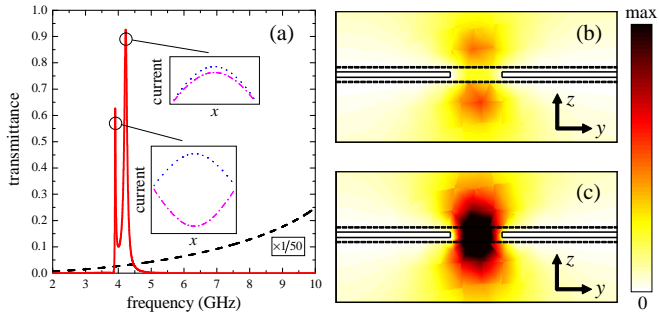}
\end{center}
\caption{(Color online) (a) Transmittance spectrum (red solid curve) at normal incidence for the geometry in Fig. \ref{Figure1} with $\Lambda_y$=30 mm, $g$=5.5 mm, $h_s$=0.5 mm, $w$=1.6 mm, $h_w$=0.2 mm, $l$=24 mm, $b$=6 mm (i.e., $\Lambda_x=l+b$=30 mm), $\Delta_x$=0, $\varepsilon_r$=3 (loss-tangent=0), and $h_d$=1.5 mm. As a reference, also shown (black dashed curve, magnified by a factor 50 for visualization purposes) is the response of the slitted screen alone. The two insets illustrate the longitudinal surface current (real part) distributions on the top (blue dotted) and bottom (magenta dash-dotted) wires, at the two resonances. (b), (c) Electric field magnitude maps (in the $z-y$ plane at the center of the wire gap, with the metallic-screen and dielectric slab regions overlaid as solid and dashed lines, respectively) at the two resonances (3.916 and 4.230 GHz, respectively).}
\label{Figure2}
\end{figure}

\section{Representative results}
Figure \ref{Figure2}(a) illustrates a representative normal-incidence transmittance ($|S_{21}|^2$) spectrum, for a configuration featuring a lossless dielectric and longitudinally symmetric [i.e., $\Delta_x=0$ in Fig. \ref{Figure1}(c)] cut-wire pairs, over a frequency range (2--10 GHz) well below the cutoff frequency of the slit (15.75 GHz, taking into account the presence of the dielectric). Two sharp resonances are clearly visible at frequencies of 3.916 GHz and 4.230 GHz, with transmittance peak values which turn out to be over three orders of magnitude (1230- and 1533-times, respectively) stronger than the response attainable in the presence of the slitted screen only (also shown as a reference). A deeper insight in the underlying mechanisms can be gained by looking at the surface current distributions on a cut-wire pair shown in the insets. In particular, the higher-frequency resonance is found to be associated with a {\em symmetric} half-wavelength current mode, which previous studies on cut-wire-pair metamaterials (see, e.g., \cite{Soukoulis2,Rhee2,Burokur}) have identified as representative of {\em electric-type} resonance phenomena. Conversely, the lower-frequency one is found to be associated to an {\em anti-symmetric} current mode, representative of {\em magnetic-type} resonance phenomena \cite{Soukoulis2,Rhee2,Burokur}. Figures \ref{Figure2}(b) and \ref{Figure2}(c) show the electric field intensity distributions in the $y-z$ plane (at the center of the wire gap), from which it is fairly evident the strong localization effect that allows the field ``squeezing'' through the slit, as also observed in \cite{Jin}. Consistently with previous observations in the literature \cite{Soukoulis2,Rhee2}, the two field distributions are rather different, with the localization most pronounced in the slit region for the electric resonance [Fig. \ref{Figure2}(c)], and around the wire ends for the magnetic resonance  [Fig. \ref{Figure2}(b)]. 

We note that the presence of two transmission resonances was also observed in \cite{Moreno1} for a configuration based only on a slit array embedded in a dielectric slab (i.e., without cut wires), and was attributed to the symmetric and anti-symmetric coupling of the surface waves supported by the slab.

%
\begin{figure}
\begin{center}
\includegraphics[width=8cm]{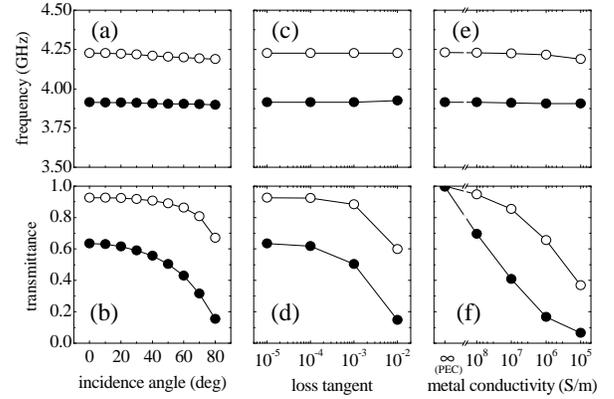}
\end{center}
\caption{Frequency and transmittance values pertaining to the electric (empty markers) and magnetic (full markers) resonances, with parameters as in Fig. \ref{Figure2}, but varying the incidence angle (a,b), the dielectric loss-tangent (c,d), and the (wire and screen) metal conductivity (e,f).}
\label{Figure3}
\end{figure}

\begin{figure}
\begin{center}
\includegraphics[width=8.8cm]{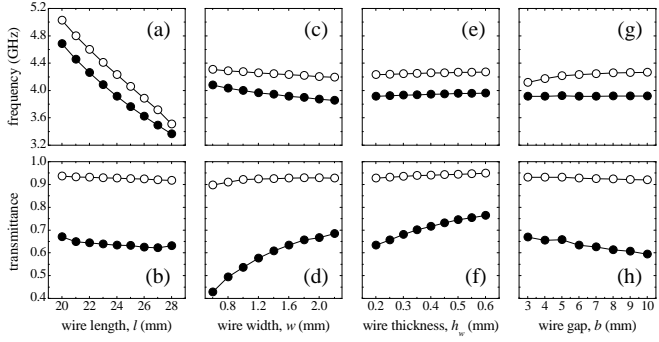}
\end{center}
\caption{As in Fig. \ref{Figure3}, but varying the wire length (a,b), width (c,d), thickness (e,f), or gap (g,h), for normal incidence and lossless dielectric.}
\label{Figure4}
\end{figure}

\begin{figure}
\begin{center}
\includegraphics[width=8.8cm]{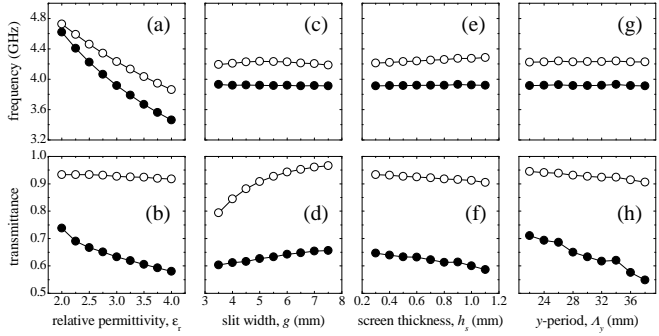}
\end{center}
\caption{As in Fig. \ref{Figure4}, but varying the dielectric relative permittivity (a,b), slit width (c,d), screen thickness (e,f), or $y-$period (g,h).}
\label{Figure5}
\end{figure}

\begin{figure}
\begin{center}
\includegraphics[width=6cm]{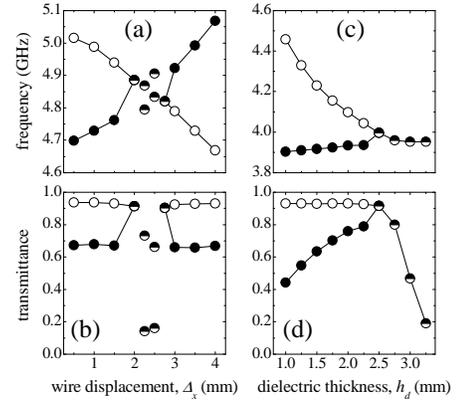}
\end{center}
\caption{As in Fig. \ref{Figure4}, but varying the wire longitudinal displacement (a,b) for $l=20$ mm, and the dielectric slab thickness (c,d). Half-empty markers indicate hybrid resonances with no clear-cut electric or magnetic nature.}
\label{Figure6}
\end{figure}

\begin{figure}
\begin{center}
\includegraphics[width=8.8cm]{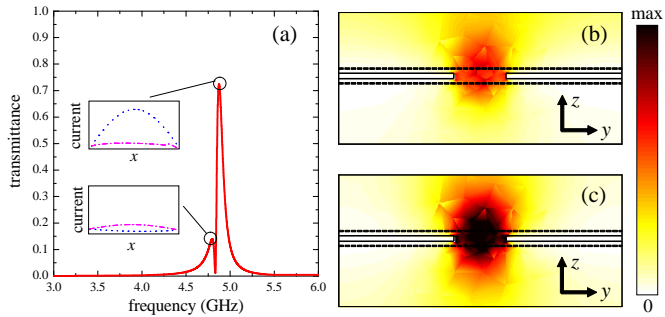}
\end{center}
\caption{(Color online) As in Fig.\ref{Figure2}, but for $l=20$ mm and $\Delta_x=2.25$ mm.}
\label{Figure7}
\end{figure}

\begin{figure}
\begin{center}
\includegraphics[width=5cm]{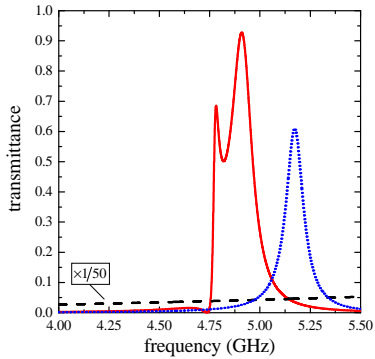}
\end{center}
\caption{(Color online) As in Fig. \ref{Figure2}(a), but for a pass-band response (with targeted transmittance $\ge 0.5$) obtained by setting the wire length $l=20$ mm and tuning the longitudinal displacement ($\Delta_x=1.75$ mm), which exhibits a relative bandwidth of nearly 4\%. Also shown (blue dotted curve), as a reference, is the response of the same structure, but with the top cut-wire array only (i.e., only one resonance, as in \cite{Jin}), which exhibits a relative bandwidth of nearly 0.9\%.}
\label{Figure8}
\end{figure}

Besides the moderately higher transmission enhancement, by comparison with \cite{Jin}, the proposed scenario features a {\em richer} phenomenology, which involves both electric- and magnetic-type resonances, typical of cut-wire-pair structures, thereby endowing further degrees of freedom in the engineering of enhanced transmission. It is therefore interesting to explore the robustness of these resonances and the possibilities of {\em tuning} them by acting on the structure parameters. In this framework, Figs. \ref{Figure3}--\ref{Figure8} compactly summarize the salient results from a comprehensive parametric study.
Specifically, Fig. \ref{Figure3} shows the resonant frequencies and transmittances observed for {\em oblique} incidence (maintaining the TE polarization), as well as in the presence of losses (up to loss-tangent=0.01) in the dielectric substrate, and for various values of the electrical conductivity of the screen and wires. Interestingly, the resonances turn out to be rather robust up to near-grazing incidence angles and moderate values of the dielectric losses, and maintain the clear-cut electric (empty markers) or magnetic (full markers) hallmarks illustrated above. A breakdown was observed for stronger loss levels (loss-tangent=0.1, not displayed), with the merging of the two resonances into a rather wide and low-transmittance peak. On the other hand, the metal conductivity appears to yield stronger effects in the peak transmittance, going from a practically {\em unit} value (for both electric and magnetic resonances) in the ideal perfectly electric conducting (PEC) case, to values over twice smaller for the magnetic resonance in the presence of {\em finite} but relatively high conductivity values ($\sigma=10^7$ S/m).

%

%

Figure \ref{Figure4} illustrates the effects induced by varying a single wire parameter (length $l$, width $w$, thickness $h_w$, or longitudinal gap $b$) with respect to the reference configuration in Fig. \ref{Figure2}. Within the parametric ranges considered, the response is still characterized by two clear-cut electric- and magnetic-type resonances, with the frequency and transmittance values pertaining to the magnetic resonances always lower than their electric counterparts. Moreover, it can be observed that both resonant frequencies are mostly sensitive to the wire length [Fig. \ref{Figure4}(a)], and very weakly dependent on the other parameters [Figs. \ref{Figure4}(c), \ref{Figure4}(e), and \ref{Figure4}(g)]. In connection with the transmittance values [Figs. \ref{Figure4}(b), \ref{Figure4}(d), \ref{Figure4}(f), and \ref{Figure4}(h)], little variations are observed for the electric resonance, while a moderate sensitivity (especially to the wire width and thickness) is observed for the magnetic one.

Qualitatively similar considerations can be drawn from Fig. \ref{Figure5}, which illustrates the effects of varying some other structural parameters (dielectric relative permittivity $\varepsilon_r$, slit width $g$, screen thickness $h_s$, and $y-$period $\Lambda_y$). In this case, the only parameter that turns out to yield sensible variations in the resonant frequencies is the dielectric relative permittivity [Fig. \ref{Figure5}(a)], while all parameters seem to affect (in mild to moderate forms) the transmittance values [Figs. \ref{Figure5}(b), \ref{Figure5}(d), \ref{Figure5}(f), and \ref{Figure5}(h)].
More complex are the effects induced by varying the wire longitudinal displacement $\Delta_x$ and the dielectric slab thickness $h_d$, as illustrated in Fig. \ref{Figure6}. In these cases, the clear-cut electric or magnetic hallmarks are not always preserved, and some {\em hybrid} forms (indicated as half-empty markers) show up in certain parametric ranges. In particular, by increasing the dielectric slab thickness (and, hence, the electrical spacing between the arrays), the two resonances tend to {\em merge} into a hybrid form in which only the top wire is significantly excited, with progressively lower transmittance values [Figs. \ref{Figure6}(c) and \ref{Figure6}(d)]. In such regime, the behavior is not very different from the {\em single} wire structure in \cite{Jin}, and the progressively lower transmittance values can be easily explained taking into account that increasing the slab thickness results in a weaker evanescent coupling. 

Conversely, by varying the wire longitudinal displacement  \cite{Burokur}, an interesting {\em crossing} phenomenon is observed [Figs. \ref{Figure6}(a) and \ref{Figure6}(b)], i.e., a {\em merging} followed by an inversion in the frequency locations of the electric and magnetic resonances, with a transition region characterized by hybrid resonances (some of which with relatively low transmittances). Such behavior is generally much more complex than the one illustrated above, and also includes situations where the current amplitudes on the wire pair are moderately different (for which a prevailing electric or magnetic behavior could still be attributed). As a representative example, Fig. \ref{Figure7} shows the transmittance spectrum and resonant features pertaining to a longitudinal displacement $\Delta_x=2.25$ mm. In the transmittance response, one can still observe two peaks with significantly different amplitudes. From the current distributions on the wires (shown in the insets) and the field maps [Figs. \ref{Figure7}(b) and \ref{Figure7}(c)], it can be observed that the lower-amplitude resonance is associated with a weak (and anti-symmetric) excitation of both wires, which results in a rather inefficient coupling, whereas the much more efficient coupling of the other resonance is mainly attributable to the strong excitation of the top wire.

%

%

The above parametric studies indicate the possibility of jointly or separately tuning the two resonances by acting on the structure parameters, which opens up interesting perspectives in the engineering of enhanced transmission phenomena. As an illustrative example, we show here a test design aimed at synthesizing a {\em pass-band} response for a given transmittance level, using as a tuning parameter the longitudinal wire displacement $\Delta_x$ [cf. Fig. \ref{Figure1}(c)]. Assuming a targeted transmittance value $\ge 0.5$ (corresponding to an enhancement factor $\gtrsim 600$ with respect to the slitted screen only), and starting from the response in Figs. \ref{Figure6}(a) and \ref{Figure6}(b) (i.e., $l=20$ mm), the best tuning was obtained for $\Delta_x=1.75$ mm, and the corresponding response is shown in Fig. \ref{Figure8}. As it can be observed, such {\em doubly-tuned} design exhibits a relative bandwidth ($\sim 4$\%) over four times larger than typical values attainable via a {\em singly-tuned} design based on the same structure featuring only the top cut-wire array (i.e., only one resonance, as in \cite{Jin}), whose typical response is also shown as a reference.

%

We point out that all the above results pertain to the case of cut-wire arrays {\em transversely centered} on the slits, which provides the highest transmission enhancements. Transverse wire displacement may also be used as a design parameter, but typically results in a less effective coupling \cite{Jin}.

\section{Conclusions}
In conclusion, we have demonstrated the potentials of paired cut-wire arrays in enhancing the transmission of TE fields through subwavelength slits in a thin metallic screen. We have shown that such enhanced transmission phenomena originate from the excitation of electric- and magnetic-type resonances supported by the cut-wire pairs. The associated resonance frequencies can be tuned by acting on the structure parameters (mainly, the wire length and longitudinal displacement, and the dielectric substrate electrical thickness), and their distance can be adjusted over a moderately wide bandwidth (nearly 10\% in the examples above).
Besides the inherent basic interest in the additional mechanisms attainable, we have shown the possibility of (jointly or separately) tuning the electric- and magnetic-type resonances, with possible application to pass-band-type designs, which may open up new perspectives in the engineering of enhanced transmission phenomena.

Current and future research is aimed at the experimental validation via prototype fabrication (in printed-circuit-board technology) and characterization, as well as at the exploration of electronic tuning mechanisms. 



\end{document}